\documentclass[]{interact}

\usepackage{epstopdf}
\usepackage{subfigure}

\usepackage[numbers,sort&compress,merge]{natbib}
\bibpunct[, ]{(}{)}{,}{n}{,}{,}

\usepackage[usenames,dvipsnames,svgnames,table]{xcolor}

\theoremstyle{plain}

\theoremstyle{definition}

\theoremstyle{remark}

\begin{document}

\articletype{ARTICLE TEMPLATE}

\title{Absolute frequency measurement of the $^2$S$_{1/2} \rightarrow ^2$F$_{7/2}$ optical clock transition in $^{171}$Yb$^+$ with an uncertainty of $4\times 10^{-16}$ using a frequency link to International Atomic Time}

\author{
\name{Charles~F.~A. Baynham\textsuperscript{a,b}, Rachel~M. Godun\textsuperscript{a}\thanks{CONTACT R.~M. Godun. Email: rachel.godun@npl.co.uk}, Jonathan~M. Jones\textsuperscript{a,c}, Steven~A. King\textsuperscript{a}, Peter~B.~R. Nisbet-Jones\textsuperscript{a}, Fred Baynes\textsuperscript{a}, Antoine Rolland\textsuperscript{a}, Patrick~E.~G. Baird\textsuperscript{b}, Kai Bongs\textsuperscript{c}, Patrick Gill\textsuperscript{a} and Helen~S. Margolis\textsuperscript{a}}
\affil{\textsuperscript{a}National Physical Laboratory, Hampton Road, Teddington, TW11 0LW, UK; \textsuperscript{b}Clarendon Laboratory, University of Oxford, Parks Road, Oxford, OX1 3PU, UK; \textsuperscript{c}School of Physics and Astronomy, University of Birmingham, Edgbaston, Birmingham, B15 2TT, UK  }
}

\maketitle

\begin{abstract}
The highly forbidden $^2$S$_{1/2} \rightarrow ^2$F$_{7/2}$ electric octupole transition in $^{171}$Yb$^+$ is a potential candidate for a redefinition of the SI second.  We present a measurement of the absolute frequency of this optical transition, performed using a frequency link to International Atomic Time to provide traceability to the SI second. The $^{171}$Yb$^+$ optical frequency standard was operated for 76\% of a 25-day period, with the absolute frequency measured to be 642~121~496~772~645.14(26) Hz. The fractional uncertainty of $4.0 \times 10 ^{-16}$
is comparable to that of the best previously reported measurement, which was made by a direct comparison to local caesium primary frequency standards.
\end{abstract}


\section{Introduction}
Optical frequency standards have already demonstrated that they can outperform caesium microwave primary frequency standards by up to two orders of magnitude in both stability and accuracy~\cite{Nicholson15, Ushijima15, Schioppo17, Huntemann16, Chou10}.  A future redefinition of the SI second (SI s) in terms of an optical transition frequency is therefore anticipated, and must be consistent with the existing definition to within the uncertainty with which it is presently realized. It is therefore essential to measure the absolute frequencies of candidate optical reference transitions with the lowest possible uncertainty.

The most direct approach to measuring the absolute frequency of an optical standard is to use a femtosecond optical frequency comb to determine the ratio between its frequency and that of a local caesium primary standard, which provides a realization of the SI second~\cite{Stalnaker07,Barwood14,Godun14,Huntemann14,Grebing16,Lodewyck16,Pizzocaro17}. However an alternative means of accessing the SI second is via a frequency link to International Atomic Time (TAI)~\cite{Park13,Akamatsu14,Huang16,Dube17,Hachisu17ApplPhysB,Hachisu17OptExp}.

TAI is a virtual time scale, computed monthly by the International Bureau of Weights and Measures (BIPM) from clock data provided by approximately 80 institutes distributed around the world, with a latency of up to 45 days. It is computed at 5-day intervals and the offset of its scale interval from the SI second is published only as a monthly average in the BIPM bulletin Circular T.
To access the SI second via this monthly value ideally requires optical frequency data to be acquired continuously over the whole month in order to avoid introducing additional uncertainty into the measurement.  Long averaging times are also required to reduce the uncertainty contribution from the satellite-based time and frequency transfer techniques used to make the link to TAI.
Since optical frequency standards do not yet commonly operate continuously over periods of many days, the dead time in their operation will therefore inflate the uncertainty of an absolute frequency measurement performed in this way. However an advantage of a TAI-based measurement is that several primary frequency standards contribute, reducing the potential systematic bias.

In this paper, we report a TAI-based absolute frequency measurement of the $^2$S$_{1/2} \rightarrow ^2$F$_{7/2}$ transition in $^{171}$Yb$^+$.
The resulting fractional frequency uncertainty of $4 \times 10 ^{-16}$ is comparable to that of the best previously reported measurement of this transition frequency~\cite{Huntemann14}, which was made at the Physikalisch-Technische Bundesanstalt (PTB) in Germany, relative to local caesium primary frequency standards. To date, only two other TAI-based absolute frequency measurements have reached fractional uncertainties below $1\times 10^{-15}$\cite{Hachisu17ApplPhysB,Hachisu17OptExp}. In those cases, the low uncertainties were achieved by using an ensemble of local flywheel oscillators to reduce the uncertainty contributions arising from intermittent operation of the optical frequency standard and/or via special computations of the TAI scale interval with respect to the SI second over evaluation periods shorter than one month. In contrast, in our work we use a single local flywheel oscillator and the standard 1-month calibration of the TAI scale interval. The low uncertainty of our absolute frequency measurement originates from the much higher up-time achieved for our $^{171}$Yb$^+$ optical frequency standard, which was operational for 76\% of a 25-day period in June 2015.

\section{Experimental overview}
The National Physical Laboratory's ytterbium ion optical frequency standard is described in more detail in reference~\cite{King12}. It is based around a single ion of $^{171}$Yb$^+$, trapped and laser-cooled in an rf end-cap trap~\cite{Nisbet-Jones16}.
Narrow linewidth light at the clock transition wavelength, 467~nm, is produced by frequency doubling an infrared laser that is stabilized to a high-finesse optical reference cavity.  The ion is repeatedly probed by the 467-nm light, and the excitation probability provides a feedback signal
to lock the laser frequency to that of the $^2$S$_{1/2} \rightarrow ^2$F$_{7/2}$ electric octupole transition. The atomic transition is perturbed by its environment and also by the probe laser itself, so corrections must be made to the output frequency in order to provide the unperturbed transition frequency.  Analysis of the various contributions to the total frequency correction, with corresponding uncertainties, is presented in section~\ref{Yb+systematics}.

Traceability of our optical frequency measurement to the SI second is achieved in several stages.
In the first step, a fibre-based femtosecond optical frequency comb is used to measure the optical frequency relative to the 10~MHz output signal from a hydrogen maser --- a robust frequency standard, which runs continuously.  The maser forms the local time scale UTC(NPL) by generating a series of pulse-per-second signals from its 10 MHz output.
The measurement performed using the frequency comb thus determines the frequency ratio between the $^{171}$Yb$^+$ optical clock transition and the frequency of the local time scale, denoted here by $f(\textrm{Yb}^+)/f(\textrm{UTC(NPL)})$.

In the second step, the local time scale UTC(NPL) is compared continuously to TAI via satellite-based time and frequency transfer links.  The time offset between UTC(NPL) and Coordinated Universal Time (UTC) is computed by the BIPM at 5-day intervals and published in the monthly Circular T bulletin. The change in the time offset between the start and the end of each 5-day period reveals the mean frequency difference between UTC(NPL) and UTC over that period.
(Note that the frequency of TAI is the same as that of UTC since the two time scales differ only by an integer number of leap seconds).
In this way we can determine the mean frequency ratio between UTC(NPL) and TAI, denoted here by $f(\textrm{UTC(NPL)})/f(\textrm{TAI})$, over any measurement period whose start and end times are aligned with the time grid on which the BIPM computations are performed.

In the final step, a correction must be made to account for the fact that the scale interval of TAI during the period of the measurement is not exactly equal to the SI second on the rotating geoid~\cite{Guinot05}.  The fractional deviation, $d$, between the TAI scale interval and the SI second on the rotating geoid is estimated by the BIPM over each one-month interval of the TAI computation and is published in Circular T.  Improved estimates are then provided later, derived from the annual calculation of TT(BIPM).

\begin{figure}
\centering
\resizebox*{15cm}{!}{\includegraphics{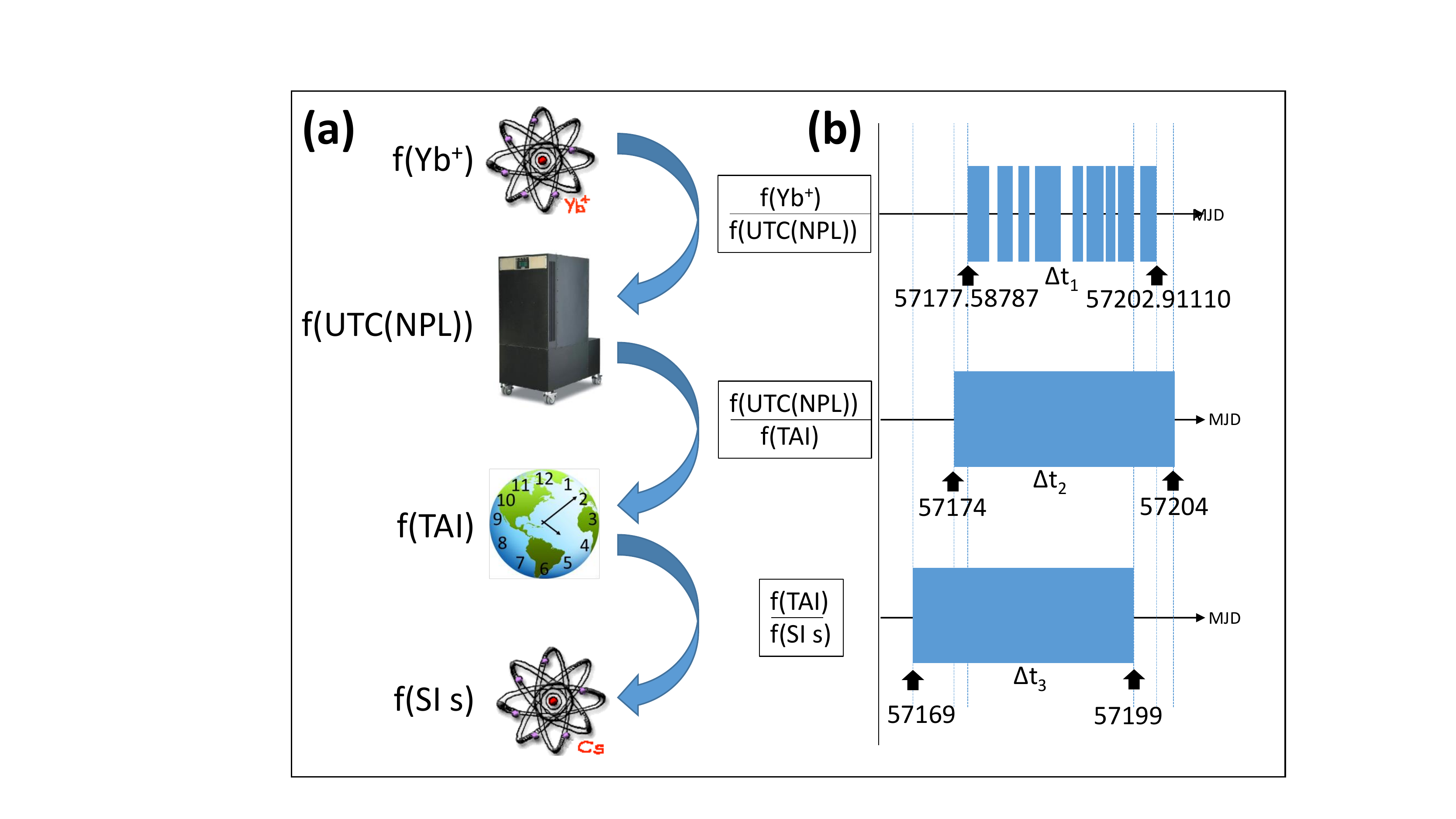}}\hspace{5pt}
\caption{(a) Schematic overview showing the chain of frequency ratio measurements used to compare the frequency of the optical transition in $^{171}$Yb$^+$ to the frequency of the microwave transition in Cs that is used to define the SI second. (b) The time intervals (blue) over which frequency ratios are measured or computed, with the start and end points indicated as modified Julian days (MJD). The time interval $\Delta t_1$ consists of a number of discontinuous intervals, whereas $\Delta t_2$ and $\Delta t_3$ are continuous.
}
\label{Fig-setup}
\end{figure}

In summary, if the optical frequency standard operates continuously throughout the one-month TAI reporting period, its absolute frequency can be evaluated as the product of three frequency ratios as depicted in Figure~\ref{Fig-setup}(a):
\begin{equation}\label{basic-ratio}
 \frac{f(\textrm{Yb}^+)}{f(\textrm{SI s})} = \frac{f(\textrm{Yb}^+)}{f(\textrm{UTC(NPL)})} \times \frac{f(\textrm{UTC(NPL)})}{f(\textrm{TAI})} \times \frac{f(\textrm{TAI})}{f(\textrm{SI s})}\,
\end{equation}
where $f(\textrm{SI s})=1\,\textrm{Hz}$ by definition.

In practice, however, there are dead times in the operation of the optical standard, and the start and end of the measurement periods do not coincide exactly with the start and end of the TAI computation period. Extrapolation of the frequency ratios is therefore necessary, and the basic formalism of Equation~\ref{basic-ratio} must be expanded to give
\begin{eqnarray}\label{full-ratio}
 \frac{f(\textrm{Yb}^+)}{f(\textrm{SI s})} & =
& \frac{f(\textrm{Yb}^+; \,\Delta t_1)}{f(\textrm{UTC(NPL)}; \,\Delta t_1)}
 \times
 \frac{f(\textrm{UTC(NPL)};\,\Delta t_1)}{f(\textrm{UTC(NPL)};\,\Delta t_2)} \nonumber \\
& \times
& \frac{f(\textrm{UTC(NPL)};\,\Delta t_2)}{f(\textrm{TAI};\,\Delta t_2)}
 \times
 \frac{f(\textrm{TAI};\,\Delta t_2)}{f(\textrm{TAI};\,\Delta t_3)} \nonumber \\
& \times
& \frac{f(\textrm{TAI};\,\Delta t_3)}{f(\textrm{SI s};\,\Delta t_3)},
\end{eqnarray}
where the time interval for the determination of each measured frequency ratio (Figure~\ref{Fig-setup}(b)) is indicated by $\Delta t_i$.  The second and fourth terms on the right hand side of Equation~\ref{full-ratio} deal with the extrapolation periods, and the associated uncertainties introduced into the absolute frequency measurement are analyzed in section~\ref{SIlinkage}.

\section{Data analysis}
To evaluate the frequency ratio, ${f(\textrm{Yb}^+)}/{f(\textrm{SI s})}$, each ratio $r=f(a)/f(b)$ on the right hand side of Equation~\ref{full-ratio} is evaluated and expressed relative to a reference ratio $r_0=f_0(a)/f_0(b)$ (Table~\ref{error-budget}). The choice of these reference ratios is arbitrary, but we choose them such that the fractional corrections $(r/r_0)-1$ have magnitudes less than $10^{-14}$. This means that the total fractional correction to the frequency ratio ${f(\textrm{Yb}^+)}/{f(\textrm{SI s})}$ can be obtained to a sufficiently high accuracy in a straightforward way by summing the fractional corrections to the individual frequency ratios.

\begin{table}
\tbl{Values ($r$) and uncertainties ($u$) of the five frequency ratios
used to determine $f(\textrm{Yb}^+)/f(\textrm{SI s})$ according to
Equation~\ref{full-ratio}.
For the ratio $f(\textrm{Yb$^+$})/f(\textrm{UTC(NPL)})$ we (arbitrarily) choose a value of $r_0$ based on the
2013 CIPM recommended frequency value of the $^{171}$Yb$^+$ optical clock transition.}
{\begin{tabular}{clccc} \toprule
 Ratio 	& Contribution 	& $r_0$ & $\left[(r/r_0)-1\right]$ 		& $u\left[(r/r_0)-1\right]$ \\
  		&				& 		&$/ 10^{-18}$	&$/ 10^{-18}$ \\
 \midrule
$\frac{f(\textrm{Yb}^+;\,\Delta t_1)}{f(\textrm{UTC(NPL)};\,\Delta t_1)}$ 									&Ratio at comb		& 642\,121\,496\,772\,645.6	& 9108 	& 100 \\
																											&Yb$^+$ statistics	& 						& 0 	& 16 \\
        																									&Yb$^+$ systematic correction	&						& -911	& 108 \\
 \midrule
$\frac{f(\textrm{UTC(NPL)};\,\Delta t_1)}{f(\textrm{UTC(NPL)};\,\Delta t_2)}$									&H-maser drift		& 1						& -312	& 20 \\	
																											&H-maser extrapolation&						& 0		& 120 \\
 \midrule
$\frac{f(\textrm{UTC(NPL)};\,\Delta t_2)}{f(\textrm{TAI};\,\Delta t_2)}$										&H-maser offset from TAI			& 1						& -7793	& 164 \\
 \midrule
$\frac{f(\textrm{TAI};\,\Delta t_2)}{f(\textrm{TAI};\,\Delta t_3)}$ 											&EAL extrapolation	& 1						& 0	&  250\\
 \midrule
$\frac{f(\textrm{TAI};\,\Delta t_3)}{f(\textrm{SI  s};\,\Delta t_3)}$												&TAI offset from SI second			& 1						& -810	& 180 \\
 \midrule
$\frac{f(\textrm{Yb}^+)}{f(\textrm{SI s})}$	&Total					&			642\,121\,496\,772\,645.6					&-718	&398 \\
 \bottomrule
\end{tabular}}
\label{error-budget}
\end{table}

\subsection{Frequency correction of the optical standard}\label{Yb+systematics}
Comparisons between NPL's femtosecond optical frequency combs have shown that they themselves introduce negligible uncertainty in an optical-microwave frequency comparison~\cite{Johnson15}. The main source of uncertainty in such a measurement in fact comes from potential frequency offsets that may arise as the 10~MHz signal from the hydrogen maser used to generate UTC(NPL) is distributed between laboratories and used to synthesize a higher frequency (8~GHz) reference against which the repetition rate of the femtosecond comb is measured. This rf distribution and synthesis is estimated to contribute an uncertainty of 1 part in $10^{16}$ to the frequency ratio measurement.

The Yb$^+$ optical frequency standard itself runs at a value that is offset from the unperturbed atomic transition frequency due to the ion's interaction with its environment.  This offset must be carefully corrected, and more details of how each of the contributing systematic frequency shifts is assessed can be found in references~\cite{Godun14, King12}. The largest perturbation comes from the ac Stark shift of the relatively high intensity probe laser that is needed to drive the nanohertz linewidth electric octupole transition $^2$S$_{1/2} \rightarrow ^2$F$_{7/2}$.  The major part of this ac Stark shift is removed in real time by using two interleaved servos, which lock the laser frequency to that of the clock transition, with a different power level in the probe pulse for each servo. The measured frequencies from the two servos are then extrapolated to zero power based on the nominal power ratio. The two probe laser powers are servo-controlled to pre-set reference levels using a photodiode placed immediately after the ion, and a separate, calibrated,
out-of-loop photodiode is used to monitor the actual powers delivered to the ion. Any residual systematic error in the ac Stark shift extrapolation,
arising from an offset between the nominal and actual power ratios,
can then be corrected in post-processing. In this way uncertainties in the extrapolated frequency can reach the parts in $10^{18}$ level. Unfortunately, however, a hardware fault developed during this measurement campaign, such that the servo-controlled powers delivered to the ion were not independently monitored at all times. As a result, the periods of data for which the power was not monitored had to be assigned a much more conservative
uncertainty, with the result that the overall fractional uncertainty contribution from the ac Stark shift was $1.06 \times 10^{-16}$. This dominates the total $^{171}$Yb$^+$ systematic uncertainty of $1.08 \times 10^{-16}$ arising from all the environmental perturbations combined.

Apart from the ac Stark shift, which was largely corrected in real-time, all other frequency offsets were corrected in post-processing in a similar manner to that presented in reference~\cite{Godun14}. The electric quadrupole shift, however, was evaluated differently.  Previously, the octupole transition frequency was measured in each of three mutually orthogonal magnetic field directions in order to average away the tensor shift that arises from the interaction between any stray electric field gradients and the electric quadrupole moment of the ion's excited state.  In this work, however, the octupole transition frequency was measured in a single magnetic field direction, and the quadrupole shift was evaluated separately by probing the ytterbium ion electric quadrupole clock transition ($^2$S$_{1/2} \rightarrow ^2$D$_{3/2}$) in three orthogonal magnetic field directions.  The quadrupole transition has an approximately fifty times larger electric quadrupole moment than the octupole transition~\cite{Schneider05,Huntemann12}
meaning that the quadrupole shift can more easily be resolved on the frequency of the quadrupole transition and hence can be used to pre-calibrate the shift on the octupole transition.  Even so, the measured shift was still indistinguishable from zero, at an uncertainty level of $1.5 \times 10^{-17}$ when scaled to the octupole transition.

The frequency shift from the blackbody radiation in the ion's environment was calculated using an improved measurement of the differential scalar polarizability of the octupole transition in
$^{171}$Yb$^+$~\cite{Huntemann16}.  The blackbody radiation shift therefore contributes only $2 \times 10^{-18}$ to the fractional uncertainty in this absolute frequency measurement.

In our previous absolute frequency measurement~\cite{Godun14},
a simple determination of the height difference between the
optical standard and the local caesium fountain was sufficient
to correct for the
gravitational redshift arising from the
gravity potential difference
between the two. In this work, where we are using TAI to provide traceability to the SI second, the gravity potential difference relative to the geoid must be determined. For this we use a value derived through measurements and computations performed as part of the International Timescales with Optical Clocks (ITOC) project~\cite{Margolis13,Denker17}.
The correction for the gravitational redshift is $-1.190(4) \times 10^{-15}$ and is included in the total fractional correction of $-9.11(1.08) \times 10^{-16}$ arising from all the frequency offsets on the ion combined.

\subsection{Frequency correction in the link to the SI second}\label{SIlinkage}
The frequency ratio
${f(\textrm{Yb}^+)}/{f(\textrm{UTC(NPL)})}$
was measured for 76\% of the period
MJD 57177.58787 - 57202.91110, with the sum of these non-continuous measurement periods denoted here as $\Delta t_1$. However due to the 5-day reporting interval in Circular T, the ratio $f(\textrm{UTC(NPL)})/f(\textrm{TAI})$ is available for a different interval,
$\Delta t_2$ (MJD 57174 - 57204). UTC(NPL) must therefore be extrapolated over the dead times in the optical data.
Since the centres of the two measurement periods
$\Delta t_1$ and $\Delta t_2$ do not coincide, a frequency correction must be applied to account for the long-term frequency drift of the maser used to generate UTC(NPL).

\begin{figure}
\centering
\resizebox*{10cm}{!}{\includegraphics{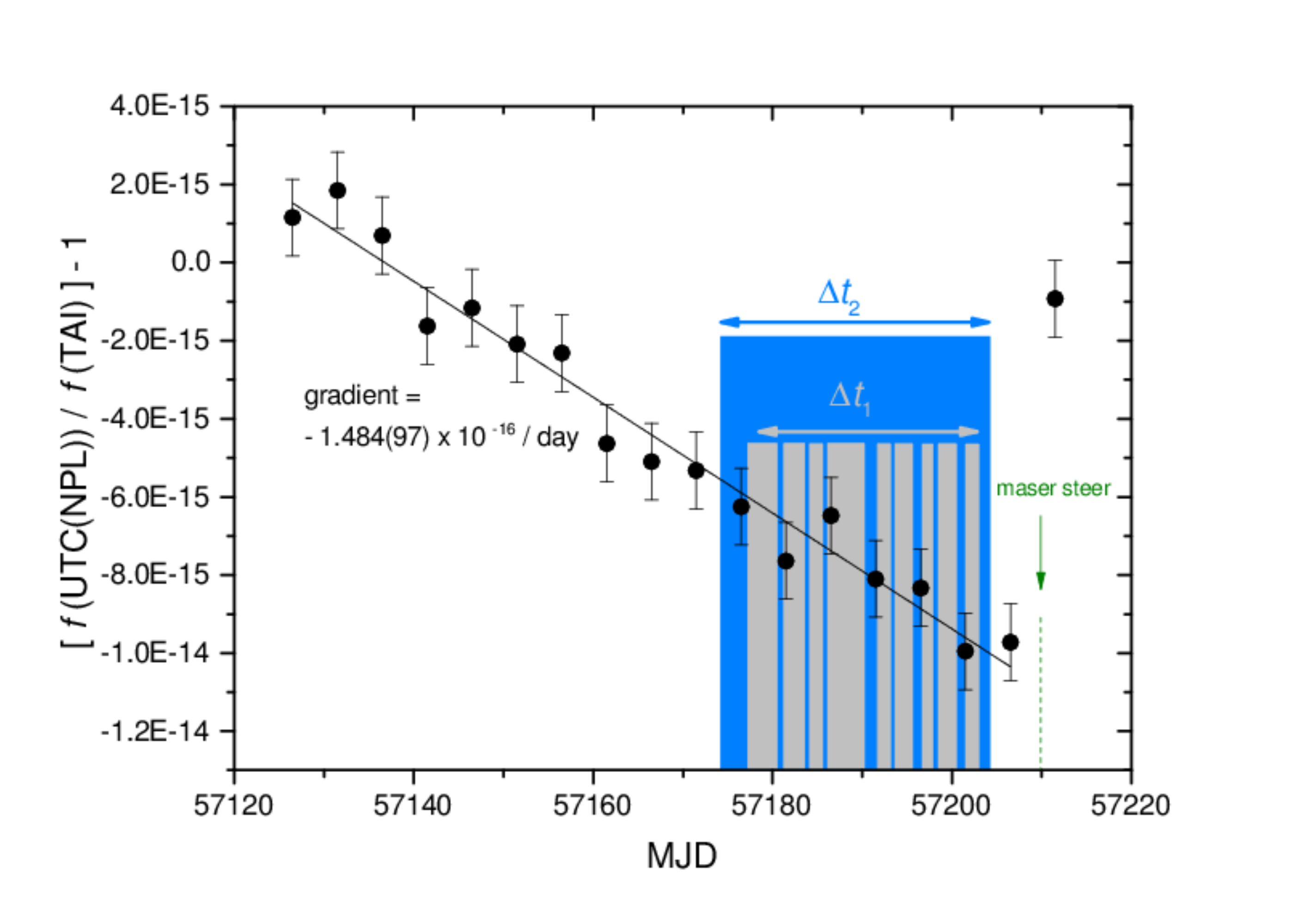}}\hspace{5pt}
\caption{Mean fractional frequency deviations of $f(\textrm{UTC(NPL)})$ from $f(\textrm{TAI})$ over successive 5-day intervals, calculated from data published in Circular T.
The frequency drift of the maser used to generate UTC(NPL) is known to be predominantly linear and so a linear fit is made to the data prior to MJD 57210, when a frequency steer was applied to the maser.
The time intervals $\Delta t_1$ and $\Delta t_2$ correspond to those in Figure~\ref{basic-ratio} as the time periods for which data was obtained for f(Yb$^+$)/f(UTC(NPL)) and f(UTC(NPL))/f(TAI) respectively.
}
\label{maser-drift}
\end{figure}

The maser's frequency drift is determined from
the values of UTC$-$UTC(NPL) provided in
section 1 of Circular T.
These values are differenced to give the mean fractional frequency offset of the maser from TAI over each 5-day period (Figure~\ref{maser-drift}) and a least-squares fit to the data from
MJD 57124 - 57209 reveals the maser's fractional frequency drift to be $-1.484(97) \times 10^{-16}/\textrm{day}$.  Since the
${f(\textrm{Yb}^+)}/{f(\textrm{UTC(NPL)})}$
data are centred around MJD = 57191.10429, whereas the period $\Delta t_2$ is centred at MJD 57189, a fractional frequency correction of $-3.12(20) \times 10^{-16}$ must be applied to account for the maser drift.

The uncertainty associated with extrapolating $f(\textrm{UTC(NPL)})$ over dead times in the operation of the $^{171}$Yb$^+$ optical frequency standard is estimated by numerical simulation, following the method outlined in~\cite{Yu07} and also used in~\cite{Hachisu15,Leute16}.  The frequency noise characteristics of the NPL
hydrogen maser, as determined from the frequency comparison against the $^{171}$Yb$^+$ optical standard, are modelled with the following fractional contributions summed in quadrature,
with $\tau$ in seconds: (i) white phase noise $4 \times 10^{-13}\,\tau^{-1}$, (ii) white frequency noise $6 \times 10^{-14}\,\tau^{-1/2}$ and (iii) flicker frequency noise $8 \times 10^{-16}$.
Two hundred data sets of simulated frequency noise,
representative of the hydrogen maser's noise processes,
were generated using Stable32~\cite{Stable32}. The difference between the average frequency
for the complete measurement period $\Delta t_2$ and for the actual measurement times $\Delta t_1$ was calculated for each data set. From the standard deviation of these frequency differences we estimate the fractional uncertainty associated with the extrapolation of $f(\textrm{UTC(NPL)})$ to be 1.20 $\times 10^{-16}$.

The mean frequency difference between UTC(NPL) and TAI during the period $\Delta t_2$ is readily obtained from the published numbers in section 1 of Circular T, yielding a fractional offset of $[ f(\textrm{UTC(NPL)}) / f(\textrm{TAI}) - 1 ]$ equal to $-7.793(164) \times 10^{-15}$.

For the fractional frequency offset $d$ between the scale interval of TAI and the SI second on the rotating geoid, we employ the BIPM computation TT(BIPM15).
For the period $\Delta t_3$ (MJD 57169 -- 57199),
$d=0.81(18) \times 10^{-15}$~\cite{SITAI15}. This 30-day evaluation interval $\Delta t_3$ is offset from the 30-day period $\Delta t_2$ by five days, as shown in Figure~\ref{Fig-setup}. The uncertainty arising from extrapolating
the frequency of TAI from period $\Delta t_2$ to $\Delta t_3$
is estimated by numerical simulation in a similar way as for $f(\textrm{UTC(NPL)})$. In this case we use
the noise characteristics of free atomic time (EAL)~\cite{Guinot05} which are stated in~\cite{SITAI15} for 2015 to be a quadratic sum of three components, with $\tau$ in days: (i) white frequency noise $1.4 \times 10^{-15}\tau^{-1/2}$, (ii) flicker frequency noise $0.3 \times 10^{-15}$ and (iii) random walk frequency noise $0.2 \times 10^{-16}\tau^{1/2}$.  Two hundred data sets of simulated frequency noise, each covering a 35-day period, were generated with Stable32~\cite{Stable32}, and the difference between the average frequency
over the first 30 days
($\Delta t_3$) and the
last 30 days
($\Delta t_2$) was calculated for each data set. The standard deviation of these frequency differences was
used to estimate a contribution of $2.50 \times 10^{-16}$ to the fractional frequency uncertainty from this extrapolation.

\section{Results and Conclusions}
\begin{figure}
\centering
\resizebox*{10cm}{!}{\includegraphics{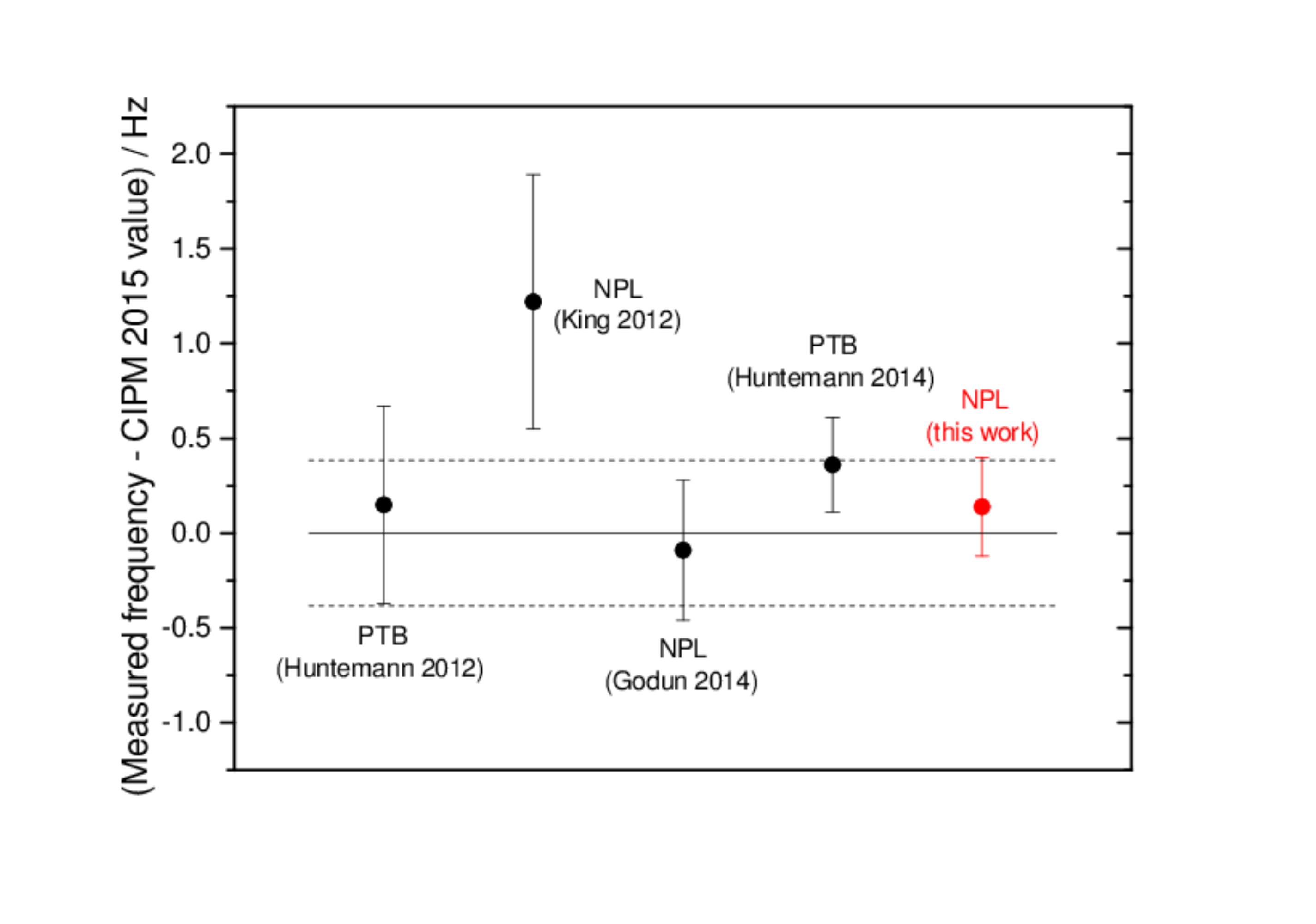}}\hspace{5pt}
\caption{Recent absolute frequency
measurements of the $^2$S$_{1/2} \rightarrow ^2$F$_{7/2}$ transition in $^{171}$Yb$^+$.  The solid and dashed lines represent the CIPM 2015 recommended value
and uncertainty.
}
\label{Yb+history}
\end{figure}

Applying all the fractional frequency corrections listed in Table~\ref{error-budget} and summing the uncertainty contributions in quadrature leads to
an absolute frequency of 642~121~496~772~645.14(26) Hz for the $^2$S$_{1/2} \rightarrow ^2$F$_{7/2}$ transition in $^{171}$Yb$^+$.
Our result
is in excellent agreement with
other
recent measurements of this transition frequency (Figure~\ref{Yb+history}), and its
fractional uncertainty of $4.0 \times 10 ^{-16}$
is similar to that of
the best published measurement
to date~\cite{Huntemann14}, which was made against local caesium fountain primary standards
rather than by using a frequency link to TAI.

Due to the very high up-time achieved for the $^{171}$Yb$^+$ optical frequency standard (76\% over 25 days), the uncertainty arising from the
extrapolation of the
local maser reference
frequency over dead times in the optical
frequency measurement
data
contributes only $1.2\times 10^{-16}$ to the overall uncertainty.
In fact the
leading contribution to the uncertainty
of our measurement
comes from the need to extrapolate the frequency of TAI from the 30-day period relevant for
the comparison of the $^{171}$Yb$^+$ optical frequency standard against
UTC(NPL),
to the 30-day period for which the frequency offset between TAI and the SI second is reported.  These 30-day periods are offset by 5 days.
In future measurements, this uncertainty contribution could be eliminated by aligning the measurement period with the reporting period of Circular T.

It is worth noting that an alternative approach to analyzing the data presented in this paper could have been taken by choosing the analysis period $\Delta t_2$ to match the 30-day reporting period for the frequency offset between TAI and the SI second ($\Delta t_3$).  This would have eliminated the need for any extrapolation of EAL, but instead would have forced greater extrapolation of UTC(NPL) between the periods $\Delta t_1$ and $\Delta t_2$.  Since UTC(NPL) is not as stable as EAL, which is formed from an ensemble of many atomic clocks, this alternative approach would have led to a larger uncertainty in the final result.

Our new frequency measurement of the $^2$S$_{1/2} \rightarrow ^2$F$_{7/2}$ transition in $^{171}$Yb$^+$ is expected to contribute to the next update of the list of recommended frequency values~\cite{MeP} maintained by the Frequency Standards Working Group (WGFS) of the Consultative Committee for Length (CCL) and Consultative Committee for Time and Frequency (CCTF). The WGFS assigns frequency values and uncertainties by performing a least-squares analysis~\cite{Margolis15,Margolis16} on a data set consisting of absolute frequency measurements and frequency ratio measurements performed by laboratories around the world. In this least-squares analysis procedure, care must be taken to properly account for any correlations between the input data, otherwise the calculated frequency values may be biassed and their uncertainties underestimated. In this context we point out that, because its traceability to the SI second is derived from TAI, the frequency measurement reported in this paper will be correlated at some level with measurements performed in other laboratories during the same period.

In fact an unusually large number of optical frequency measurements, as well as several frequency ratio measurements, were performed during June 2015. This was a result of a coordinated campaign to compare optical atomic clocks and caesium fountains in four European laboratories via satellite links, which was performed as part of the ITOC project~\cite{Margolis13}. As a result, our absolute frequency measurement is, for example, correlated with absolute frequency measurements of $^{87}$Sr~\cite{Lodewyck16} and $^{199}$Hg~\cite{Tyumenev16} optical standards performed at LNE-SYRTE and an absolute frequency measurement of the $^{87}$Sr reference transition performed at PTB~\cite{Grebing16}. Calculation of the relevant correlation coefficients will require the laboratories concerned to exchange detailed information about exact up-times of the optical clocks and caesium fountains, as well as a knowledge of the weighting applied to each Cs fountain in the computation of TAI by the BIPM.

As the robustness and reliability of optical frequency standards continue to improve,
up-times similar to those reported here
will become more routinely achievable. This will make it
possible to measure absolute frequencies at the low parts in $10^{16}$ level
in an increasing number of laboratories, even those
where local primary standards are not available,
and to operate the optical standards as secondary representations of the second contributing to TAI.

\section*{Acknowledgements}
We thank Peter Whibberley for helpful discussions and E. Anne Curtis for critical review of the manuscript prior to submission. We also note that our absolute frequency measurement derives its accuracy from the primary standards operated at other national measurement institutes around the world.

\section*{Funding}
This work was financially supported by the UK Department for Business, Energy and Industrial Strategy as part of the National Measurement System Programme; the European Metrology Research Programme (EMRP) project SIB55-ITOC; and the European Metrology Programme for Innovation and Research (EMPIR) project 15SIB03-OC18. This project has received funding from the EMPIR programme co-financed by the Participating States and from the European Union’s Horizon 2020 research and innovation programme. The EMRP is jointly funded by the EMRP participating countries within EURAMET and the European Union.

\end{document}